\documentclass[twocolumn]{aastex6}

\pdfoutput=1

\usepackage{amsmath}
\usepackage{amsfonts}
\usepackage{amssymb}
\usepackage{wasysym}
\usepackage[percent]{overpic}

\makeatletter
\def\env@matrix{\hskip -\arraycolsep % taken from amsmath.sty lines 895ff
  \let\@ifnextchar\new@ifnextchar
  \array{*{\c@MaxMatrixCols}c}}
\makeatother

\usepackage{graphicx}	% Including figure files
\usepackage{amsmath}	% Advanced maths commands
\usepackage{amssymb}	% Extra maths symbols
\usepackage{bm}		% Bold maths symbols, including upright Greek
\usepackage{hyperref}

\hypersetup{
    colorlinks=true,
    citecolor=black,
    urlcolor=cyan,
}
\slugcomment{Draft version}

\shorttitle{The Spaceline}
\shortauthors{Penoyre \& Sandford}

\begin{document}

\title{The Spaceline: a practical space elevator alternative achievable with current technology}

\author{Zephyr Penoyre$^{1}$ and Emily Sandford$^2$}
\affil{$^1$Institute of Astronomy, University of Cambridge, Madingley Road, Cambridge, CB3 0HA, United Kingdom}
\affil{$^2$Dept. of Astronomy, Columbia University, 550 W. 120th Street, New York NY 10027, USA}

\begin{abstract}
Perhaps the biggest hurdle to mankind's expansion throughout the Solar System is the prohibitive cost of escaping Earth's gravitational pull. In its many forms the \textit{space elevator} provides a way to circumvent this cost, allowing payloads to traverse along a cable extending from Earth to orbit. However, modern materials are not strong enough to build a cable capable of supporting its own weight. In this work we present an alternative to the classic space elevator, within reach of modern technology: The \textit{Spaceline}. By extending a line, anchored on the moon, to deep within Earth's gravity well, we can construct a stable, traversable cable allowing free movement from the vicinity of Earth to the Moon's surface. With current materials, it is feasible to build a cable extending to close to the height of geostationary orbit, allowing easy traversal and construction between the Earth and the Moon.
\end{abstract}

In preparation for submission to Acta Astronautica. Questions and suggestions welcome.
%\keywords{In preparation for submission to Acta Astronautica - questions and suggestions welcome}

\section{Introduction}
%tktk Intro to space elevators
For a vehicle travelling in empty space it's momentum, as well as it's energy, comes from it's fuel. It must push heavy material behind it to propel itself forward. However, if there were a fixed object to push against a vehicle could generate momentum via friction. Thus if we could design a steady cable, in tension, spanning a region of deep-space, we could move along it with solar power (or any other such source) alone. This can greatly decrease the cost and difficulty of spanning extraterrestrial distances - and is the reason why the concept of the space elevator (a cable, held vertical by centrifugal force, from the equator into deep space) is seen a major leap in reducing the cost, and increasing the access, of human space-travel.

However, the fundamental limit on a space elevator is whether a material can support its own weight over the necessary length. Modern, mass-producible materials currently \textit{cannot} reach this limit, either they break under their own weight or must be so wide as to be near implausible to construct and deploy. We may be on the horizon of substantially stronger materials, such as carbon nanotubes, but the maximum manufacturable length of such materials is presently prohibitively short.%tktk CITE.

However, there is another way to circumvent this problem, and that is to reduce the force on the cable. Classic space elevators are supported by centrifugal forces, and thus require a large counterweight beyond geostationary orbit to counteract Earth's gravitational pull. These competing forces put the cable in a large amount of tension ,and even when the cable has a tapered profile (which minimises the tension), there is a large weight of cable close to the Earth which experiences huge gravitational forces.

In comparison, a cable which only hangs into Earth's gravitational well need not be thick or massive. It is optimal to make it as thin as possible as it extends closer to Earth. This means that the gravitational forces the cable feels, and thus the tension, is much reduced.

\begin{figure}[ht]
\includegraphics[width=0.95\columnwidth]{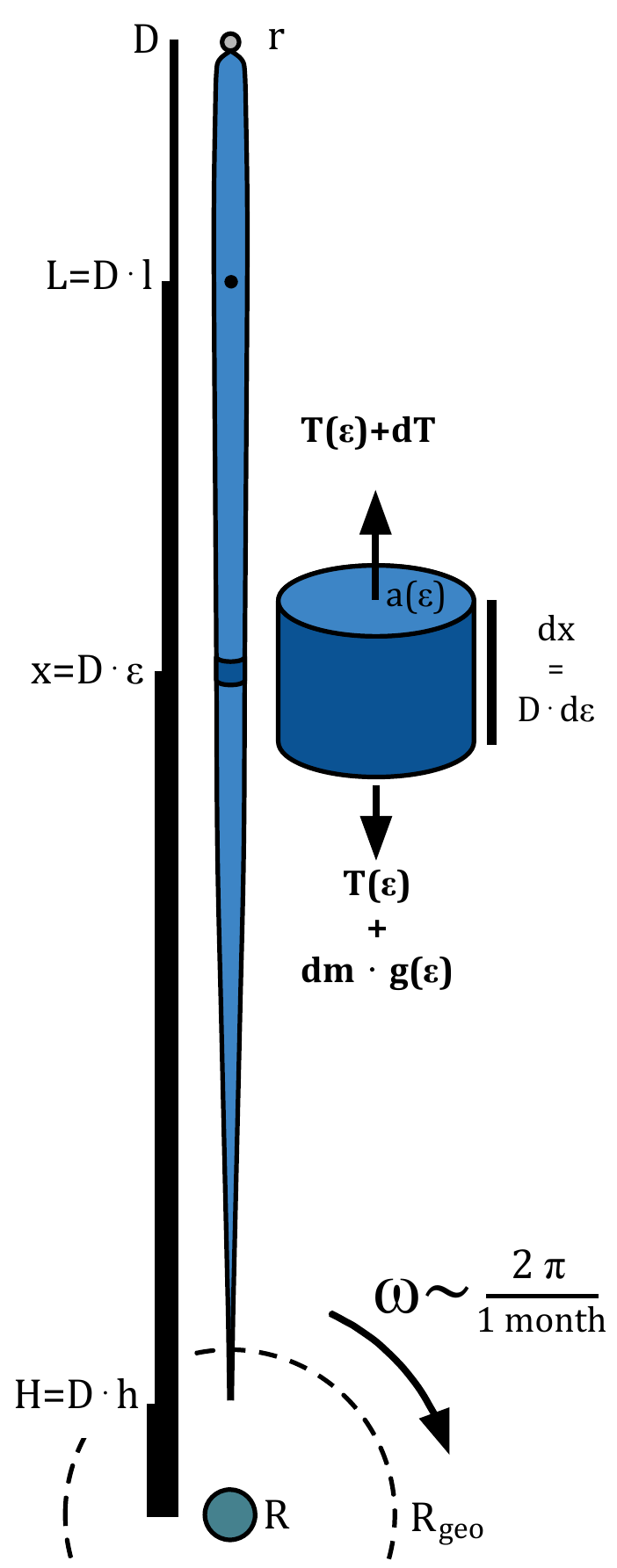}
\caption{A simple sketch of the Spaceline, with the Earth (radius $R$), Moon (radius $r$), Earth-Moon distance ($D$), and height of geostationary orbit ($R_{\mathrm{geo}}$) drawn to scale. Throughout this work, we will rescale the physical coordinates, $x$, to dimensionless coordinates, $\epsilon$, such that the centre of the Earth is at $\epsilon=0$ and the centre of the Moon at $\epsilon=1$.}
\label{scale}
\end{figure}

This is the basic concept of what we dub \textit{the Spaceline}. Figure~\ref{scale} shows a conceptual sketch of the spaceline. We will show that with materials and techniques comparable to what has already been achieved in the fields of manufacturing and spaceflight, such a cable, extending to the height of a geostationary orbit, is theoretically achievable today. Once the cable is constructed, the cost of subsequent spacelines will diminish dramatically, and correspondingly greater payloads will be transportable, relatively cheaply, between geostationary orbit and the Moon.

This is not a completely novel concept, rather an independent genesis and derivation of an idea that has been explored in works like \citet{Pearson79}, \citet{Pearson05} and \citet{Eubanks16}, where it has also been called a \textit{Lunar Space Elevator}\footnote{A name we avoid so as to highlight the different forces acting upon the cable - namely that the Spaceline does not need centrifugal forces to be in equilibrium}. We present the derivations herein as a full standalone mathematical and physical description of the concept, one that we and authors before us have been surprised to find is eminently plausible and may have been overlooked as a major step in the development of our capacity as a species to move within our solar system.

\begin{table*}
\centering
\begin{tabular}{ c c | c} 
 $M$ & Mass of Earth & $5.972 \times 10^{24}$ kg  \\
 $m$ & Mass of Moon & $7.348 \times 10^{22}$ kg $\approx$ $0.01 \  M$ \\
 \hline
 $R$ & Radius of Earth & $6,371$ km  \\
 $r$ & Radius of Moon & $1,737$ km $\approx$ $0.27 \ R$ \\
 $D$ & Earth-Moon distance & $384,400$ km $\approx$ $60 \ R$ \\
 \hline
 $R_{\mathrm{ISS}}$ & International Space Station altitude & $\sim 100$ km above Earth's surface $\approx 6,771$ km $\approx$ $1.06 \ R$ \\
 $R_{\mathrm{geo}}$ & Geostationary orbit & $42,164$ km $\approx$ $6.6 \ R$ \\
 $L$ & Earth-Moon Lagrange point 1 & $326,000$ km $\approx$ $51 \ R$
 \\
\end{tabular}
\caption{The parameters of the Earth-Moon system. The Earth-Moon distance given is the semi-major axis of the Moon's orbit, which is slightly eccentric ($e=0.055$). Currently, we do not account for the Moon's eccentricity, but it represents only a small correction to our calculations.}%tktk but we discuss it further in Correcting for this is a small change to the calculations which, for brevity, we do not make in this work, but do discuss further in section \ref{practical}.}
\label{earthMoon}
\end{table*}

\section{Preliminaries}

Before we introduce the construction of the Spaceline itself, it will be useful to define the problem in simple quantities. 

\subsection{Forces in the Earth-Moon system}

Figure~\ref{scale} shows a simple sketch of the Earth-Moon system, and Table~\ref{earthMoon} gives some relevant physical scales. It will be convenient to work in dimensionless co-ordinates, scaled relative to the Earth-Moon distance $D$. If $x$ is the distance from the centre of the Earth to some point, let
\begin{equation}
\epsilon = \frac{x}{D}.
\end{equation}
Thus, a cable extending from the surface of the moon to some height $H$ above the Earth, would run from $\epsilon = 1-\frac{r}{D}$ to $h$ (where $H=hD$) in our co-ordinates. For reference, geostationary orbit occurs at $\epsilon_{geo} \approx 0.12$ and the Lagrange point between the Earth and the Moon, where the gravitational forces of Earth and the Moon balance exactly, occurs at $l \approx 0.85$.

The Spaceline is held in place primarily by its gravitational attraction to the Earth. It is anchored to the Moon at one end, and the weight of the cable itself keeps it pulled taut, pointing toward Earth, by Earth's gravity. It rotates with the Moon, completing a full rotation once per lunar orbit ($\sim 1$ month). As the Moon is tidally locked, the anchor point will always face directly towards the Earth and we need not worry about the cable winding itself up.

In contrast, the classic space elevator is supported by large centrifugal forces, anchored to and co-rotating with Earth, completing a full rotation once a day. As it is fixed at one point on the Earth's surface, it must always remain vertically above that point. If there were significant differential rotation, the cable would start to bend and collapse. We save a detailed derivation of the form and forces of the space elevator for Appendix~\ref{elevator}.

We can write down the gravitational acceleration acting on a point mass a distance $x$ from Earth and distance $D-x$ from the Moon,
\begin{equation}
    a_{grav}(x) = \frac{GM}{x^2} - \frac{Gm}{(D-x)^2}.
\end{equation}
Note that we define all accelerations as pointing towards the Earth, such that they will be negative close to the Moon.

In the co-rotating Earth-Moon frame, the centre of mass is at
\begin{equation}
    x_{c.o.m.}\footnote{This actually lies within the Earth's radius and hence the assumption that the Earth can be treated as a point mass for this calculation breaks down. However, the difference will be small, and only affects the centrifugal forces which, as we will go on to show, are only a small fraction of the forces acting on the cable. Thus we will not concern ourselves with this detail presently.}=\frac{m}{m+M} D.
\end{equation}
The Earth-Moon frame rotates with an angular velocity
\begin{equation}
    \omega = \sqrt{\frac{G(M+m)}{D^3}}.
\end{equation}
Thus the centrifugal acceleration is
\begin{equation}
    a_{cent}(x) = (x_{c.o.m.}-x)\omega^2 = \frac{G(M+m)}{D^3} \left[\frac{m D}{M+m} - x \right].
\end{equation}

The stationary Spaceline feels no Coriolis forces. Payloads moving along it will, although at speeds of up to $0.01c$, the resulting bowing of the cable is negligible. 

It will also be useful to express the mass of the Moon in terms of the mass of the Earth using
\begin{equation}
    \mu = \frac{m}{M}.
\end{equation}

We can combine the gravitational and centrifugal accelerations to find the total acceleration. Expressing it simply in terms of $\epsilon$ and $\mu$ gives
\begin{equation}
g(\epsilon) = \frac{GM}{D^2} \left[\frac{1}{\epsilon^2} - \frac{\mu}{(1-\epsilon)^2} + (\mu - (1+\mu)\epsilon) \right].
\end{equation}

The first two terms in the brackets are the gravitational contribution, and it can be seen that these dominate close to the Earth and Moon respectively. The third term is the centrifugal force, which is generally small (and can be ignored for pedagogically simpler calculations with only a minor impact on the results).

\begin{table*}
\centering
\begin{tabular}{ c | c c c c} 
 Material & $\rho$ (kgm$^{-3}$) & $B$ (Nm$^{-2}$) & $S$ (Nmkg$^{-1}$) & $\alpha$ \\
 \hline
 \hline
 Steel & $8000$ & $5\times10^8$ & $6.25\times10^4$ & 0.06 \\
 Titanium alloy & $4800$ & $1.25\times10^9$ & $2.6\times10^5$ & 0.24 \\
 Spider silk & $1300$ & $1.4\times10^9$ & $1\times10^6$ & 0.98 \\
 Carbon fibre & $1750$ & $4.3\times10^9$ & $2.5\times10^6$ & 2.2 \\
 Kevlar & $1400$ & $3.6\times10^9$ & $2.5\times10^6$ & 2.5 \\
 Dyneema & $970$ & $3.6\times10^9$ & $3.7\times10^6$ & 3.4 \\
 Zylon & $1500$ & $5.8\times10^9$ & $3.9\times10^6$ & 3.5 \\
 \textit{Carbon nanotube} & $\sim 1000$ & $6\times10^{10}$ & $6\times10^7$ & 55 \\
\end{tabular}
\caption{The density ($\rho$), tensile strength/breaking stress ($B$), specific strength ($S$) and relative strength ($\alpha$) of various materials. Kevlar, Dyneema and Zylon are trademarked names for three different carbon-based manufactured fibres from different molecular families. Carbon nanotubes have been produced in laboratory conditions, but are still many years from being mass producible. Values taken straight from \href{https://www.wikiwand.com/en/Specific_strength}{Wikipedia}.}
\label{materials}
\end{table*}

\subsection{Physical constraints on a cable}

A few simple parameters, intrinsic to the material of which the cable is made, define the physical capabilities of a cable in tension. Ignoring defects and wear, a material will break when the stress (force per unit cross-sectional area $a$) exceeds some critical value, the breaking stress $B$. Thus, a heavier load and larger force can be accommodated by a cable made of stronger material (higher $B$) or a cable with larger cross-sectional area (higher $a$).

The density of the material, $\rho$, which can reasonably be assumed to be constant and intrinsic to the material, will also be important. Most of the load a cable must bear is its own weight. It is very possible to invent a construction such that a cable would break solely from the forces acting upon it, before we introduce any payload weight (in fact, this is the major stumbling point of current space elevator design). Thus it is useful to define the specific strength, $S=\frac{B}{\rho}$, such that a ``specifically stronger" material is one with either higher breaking stress or lower density.

Finally, it will be useful to reparameterise the strength of the material in terms of the relevant forces for the problem we are trying to solve. We define the dimensionless \textit{relative strength} of a material as
\begin{equation}
    \alpha = \frac{SD}{GM}.
\end{equation}
A material with $\alpha \gg 1$ will be much stronger than the forces involved in the Earth-Moon system, and materials with $\alpha \ll 1$ will be much too weak.

In Table~\ref{materials}, we list the strength of a range of materials. Man-made materials, particularly those made from long chains of carbon, can now achieve very high strengths (finally surpassing those from the natural world) without very high densities. Carbon nanotubes, which are made of graphene manipulated into tube-like structures, are included in this table as they have been synthesised, but only very short lengths have been produced. They provide a very promising glimpse to the future of strong, lightweight materials, but their use in large-scale engineering projects is still a long way off. Many variants on the space elevator invoke these future materials to make the design feasible, but as we will show, the strength of existing materials, such as Kevlar and Dyneema, is more than sufficient to construct a spaceline.

\begin{figure}[ht]
\includegraphics[width=\columnwidth]{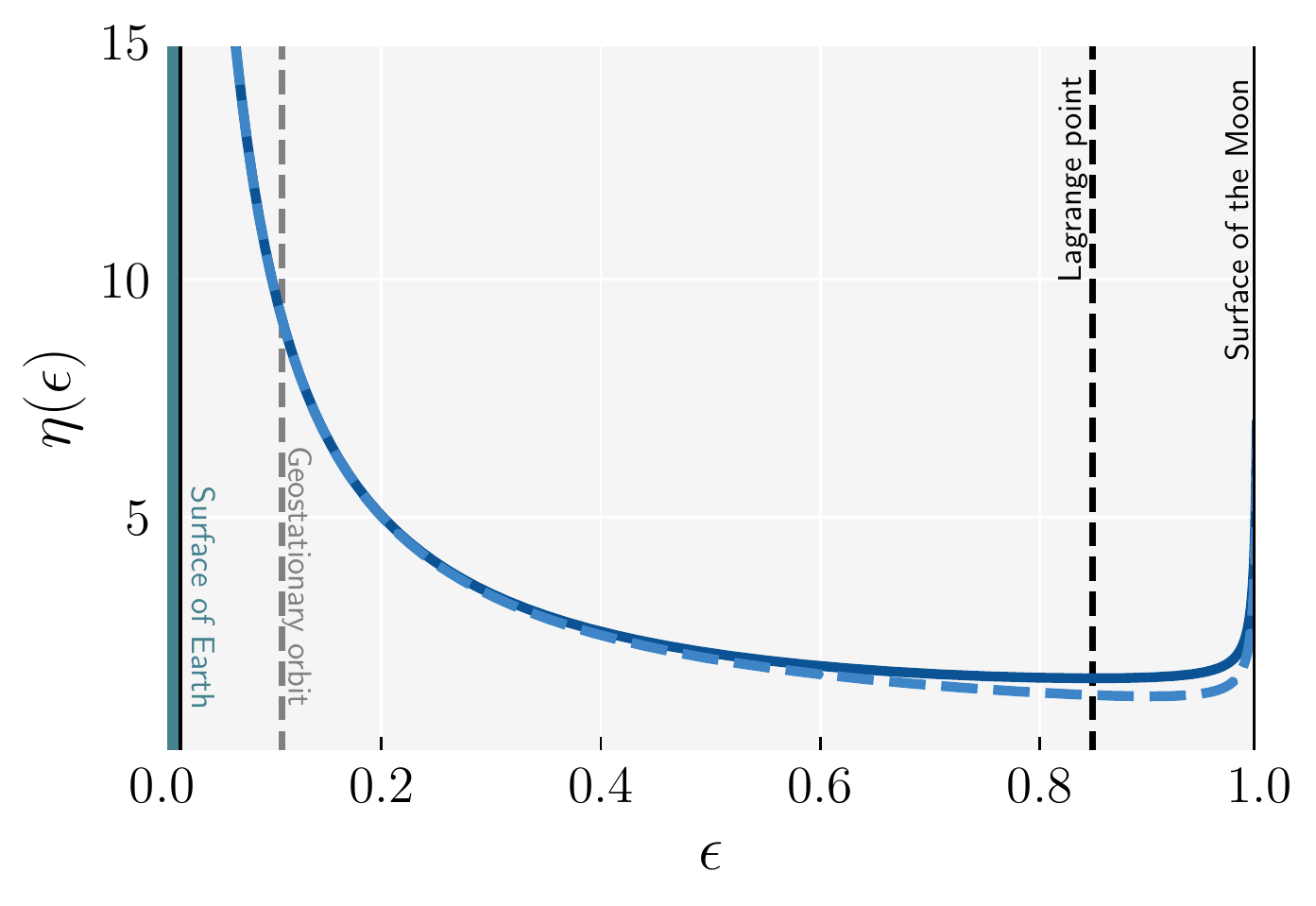}
\caption{The variation of $\eta(\epsilon)$, as given by Equation~\ref{etaEq}, as a function of the scaled distance from the Earth, $\epsilon$, which ranges from 0 at Earth's centre to 1 at the centre of the Moon. The dashed curve shows the form if centrifugal forces are ignored. The height of Earth's surface, a geostationary orbit and the Lagrange point are all shown (as is the surface of the Moon, though it is too close to the right-hand edge of the plot to be visible here).}
\label{eta}
\end{figure}

\subsection{Forces on the cable}

Figure~\ref{scale} also diagrams an infinitesimal cross-section of the cable, of length $dx$ at position $\epsilon$, and the forces acting upon it. It feels the tension of the weight of cable ``below" (closer to Earth than) it, $T(\epsilon)$; the gravitational and centrifugal forces acting upon it, $ g(\epsilon) dm$; and the tension of the cable above (Moonward of) it, $T(\epsilon+d\epsilon) = T(\epsilon)+dT$.

The mass of this infinitesimal element, with length $dx= D d\epsilon$, is
\begin{equation}
dm = \rho a(\epsilon) D d\epsilon
\end{equation}
where $a(\epsilon)$ is the cross-sectional area of the cable at $\epsilon$ (the cable's area may vary along its length). Thus, the tension of the cable satisfies
\begin{equation}
\begin{aligned}
\label{dTension}
    dT &= dm \ g(\epsilon) \\ =& \frac{GM\rho}{D} a(\epsilon) \left[\frac{1}{\epsilon^2} - \frac{\mu}{(1-\epsilon)^2} + (\mu - (1+\mu)\epsilon) \right] d\epsilon.
\end{aligned}
\end{equation}
Note that at the Lagrange point ($\epsilon = l \approx 0.85$), $dT$ will equal zero. Earthward of the Lagrange point, $dT > 0$, and Moonward of the Lagrange point, $dT < 0$; correspondingly, the tension $T$ is greatest at the Lagrange point.

If the area of the cable is specified, $dT$ can be integrated to find the tension throughout the cable. The free end of the cable, the end closest to the Earth, experiences no tension. If the cable terminates at a height $H$ (measured from the centre of the Earth) the boundary condition is $T(h)=0$ where $h=\frac{H}{D}.$

Therefore, for a given material and cross-sectional area profile $a(\epsilon)$, Equation~\ref{dTension} can be integrated to find the total tension, $T(\epsilon)$, and we can thus calculate the conditions under which a given cable will break. We save such calculations for section \ref{designs}, where we introduce the simple and optimal forms of $a(\epsilon)$.

As we have discussed, a cable will break if the stress it experiences at any point exceeds the breaking stress. It will be convenient to work in terms of the \textit{specific stress},
\begin{equation}
\sigma(\epsilon) = \frac{T(\epsilon)}{\rho a(\epsilon)}.
\end{equation}

When $\sigma(\epsilon)$ is equal to the specific strength $S$, the cable snaps.

At the same time, a cable can only support itself in tension; it has no strength in compression (unlike a tower of stacked bricks, for example, where the opposite would be true). 

Thus there are two ways in which the spaceline can fail:
\begin{itemize}
    \item Collapse - The tension, $T(\epsilon)$, must always be positive; otherwise the cable will collapse under its own weight, as it cannot support itself in compression. Since the tension decreases outward in both directions from the Lagrange point, it is minimised at the two ends of the cable. To avoid the cable collapsing onto the surface of the moon, we therefore require that the tension be positive at the moon's surface: $T(1-\frac{r}{D}) \ge 0$.
    \item Breaking - The stress must never exceed the breaking stress, or else the cable will snap. The tension (and thus the stress) is largest at the Lagrange point; thus, to avoid breaking, we require $\sigma(l)<S$.
\end{itemize}

%THIS IS NOT CURRENTLY ADDRESSED ANYWHERE
%There are a few further practicalities of such a cable, such as the addition of a payload and the stability of the system, but we will save a more detailed discussion of these for Section \ref{practical}.

\section{Spaceline designs}
\label{designs}

In this section we explore three designs of the Spaceline, all stretching from the Moon's surface to close to the Earth ($\sim R_{\mathrm{geo}}$). The difference between these designs lies only in their profile, i.e. how the cross-sectional area of the cable changes along its length.

When the cross-sectional area $a(\epsilon)$ is specified, Equation~\ref{dTension} can be integrated, and we can then see whether the conditions are met for the cable to neither break nor collapse.

Assuming the strictest constraint on building the Spaceline is the mass (and volume) of cable that needs to be transported into space, we seek the lightest possible configuration that will neither break nor collapse.

We derive the tension and stress in (i) a cable with constant cross-sectional area, (ii) one with a tapering profile, and finally (iii) a hybrid of the two. We will show that the hybrid is the most efficient, given the practicalities of constructing such a structure, but all three are relevant and help build intuition and understanding.

It will be useful to introduce the function
\begin{equation}
\begin{aligned}
\label{etaEq}
    \eta(\epsilon) =& \int - \frac{D^2}{G M} g(\epsilon) d\epsilon \\
    =& \int - \left(\frac{1}{\epsilon^2} - \frac{\mu}{(1-\epsilon)^2} + (\mu - (1 + \mu)\epsilon)\right) d\epsilon\\
    =& \frac{1}{\epsilon} + \frac{\mu}{1-\epsilon} + \frac{(1+\mu)\epsilon^2}{2} - \mu \epsilon 
\end{aligned}
\end{equation}
which will appear frequently throughout this section (the minus sign in the integrand ensures $\eta$ is always positive). For reference, Figure~\ref{eta} shows the behaviour of this function, which becomes very large close to the Earth and, to a lesser extent, near the Moon. It reaches a minimum at the Lagrange point.

\begin{figure}[ht]
\includegraphics[width=\columnwidth]{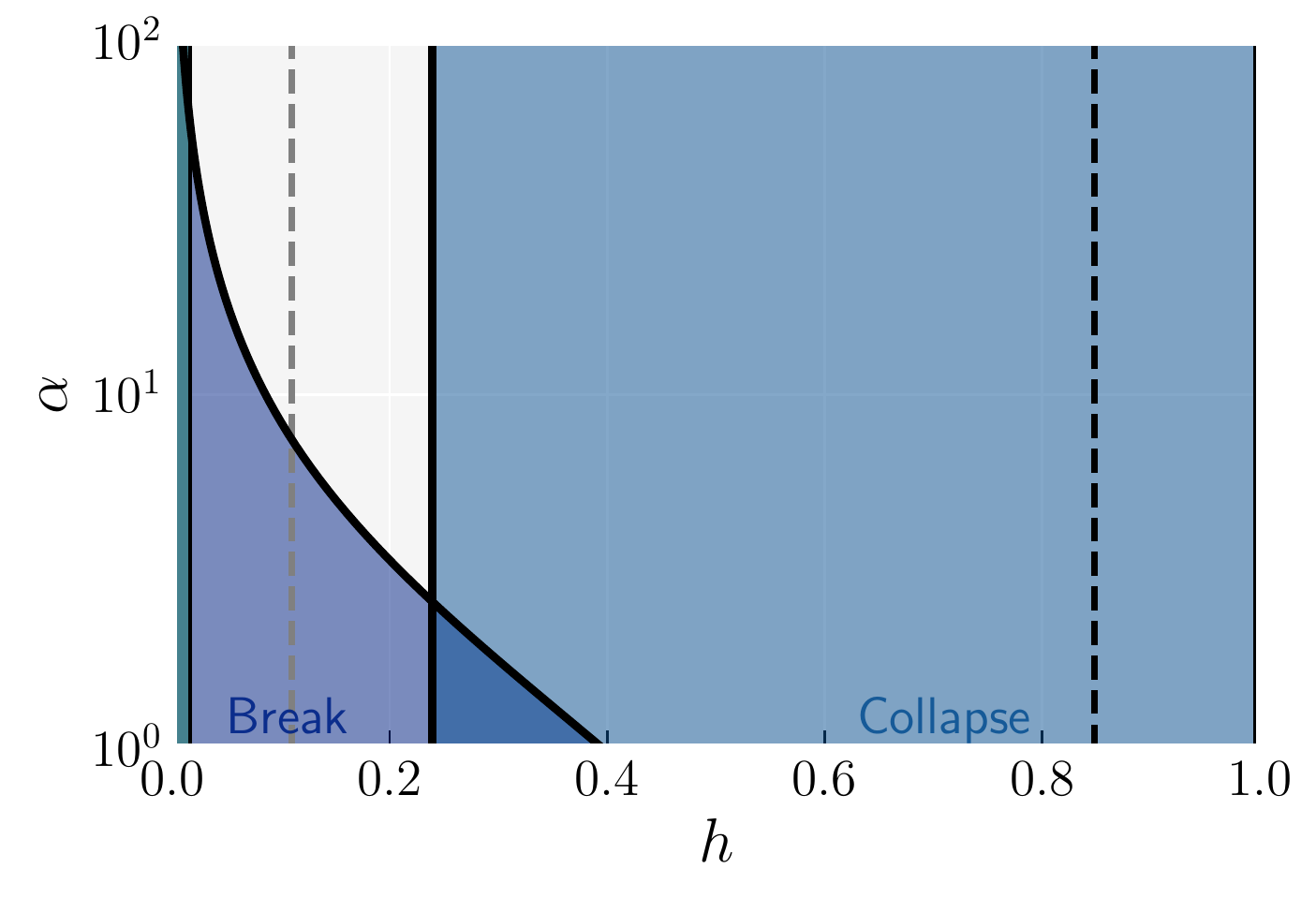}
\caption{The landscape of outcomes for the uniform-area cable, depending on the relative strength $\alpha$ and the scaled distance between Earth's centre and the free end of the cable $h$. The white region shows the possible cable heights for which a uniform-area spaceline can be constructed without breaking or collapsing, and the corresponding requirements for the material strength.}
\label{uniform}
\end{figure}

\subsection{Constant area cable}

The simplest cable is one which has a constant cross-sectional area along its entire length. Thus
\begin{equation}
a(\epsilon)=a_0
\end{equation}
and Equation~\ref{dTension} can be integrated to give
\begin{equation}
T(\epsilon) = \frac{GM\rho a_0}{D} \left[ \eta(h) - \eta(\epsilon)\right],
\label{uniformT}
\end{equation}
where $h$ is the scaled distance between Earth's centre and the free end of the cable.

Furthermore, we can find the specific stress:
\begin{equation}
\sigma(\epsilon) = S \frac{\eta(h) - \eta(\epsilon)}{\alpha}.
\end{equation}

From the tension, given by Equation~\ref{uniformT}, we can show that the cable will not collapse (i.e., $T>0$ always) if $\eta(h) > \eta(1-\frac{r}{D})$. This gives a maximum value of $h$: it must always be less than $\sim 0.24$, irrespective of the material. This requirement is not overly constraining: adding an anchorweight to the free end of the cable allows the creation of a stable structure for any $h<l$, though for large $h$ the mass of this anchorweight can become prohibitively large. However, any anchorweight must weigh more than the length of cable it replaces, and thus for an Earth-Moon cable there is no obvious utility to using one.% One could even build a modest (compared to the scales involved) compression based structure on the moon extending up to the height where $T=0$.

The condition for the cable not to break becomes the condition that $\eta(h) < \alpha + \eta(l)$ (remembering that the stress is maximized at the Lagrange point). $\eta(l)$ is a constant of the system, and is approximately equal to 1.6, and thus we see the impact of the relative strength; larger $\alpha$ allows larger values of $\eta(h)$, which correspond to smaller values of $h$, without reaching the breaking stress. This means that for larger $\alpha$, the cable can extend closer to Earth's surface.

We now have two conditions on $\eta(h)$, both of which must be satisfied for a uniform-area spaceline to be possible. We can rearrange these to a condition on the strength of the material, $\alpha > \eta(1-\frac{r}{D}) - \eta(l) \sim 2.6$. Thus a constant area spaceline can exist for any material with a relative strength of $\sim 3$ or greater, which, with reference to Table~\ref{materials}, a number of currently manufacturable materials satisfy.

This is in stark contrast to a similar calculation applied to the space elevator (see Appendix~\ref{cElevator}), which shows that the space elevator would require a material with $\alpha > 50$, impossible with modern materials.

Figure~\ref{uniform} shows how strong a cable material must be, to avoid breaking, as a function of the height of the free end of the cable, $h$. We see that as the strength increases, the free end of the cable can come closer to the Earth's surface before it breaks. A cable with $\alpha > 50$ would be able to reach Earth's atmosphere/surface (although of course such a strong material could also allow the construction of the space elevator).%; see Section~\ref{practical} for a more in depth discussion of what this would mean for space-faring.

\subsection{Tapering cable}

If a constant-area cable breaks, it does so at one specific point where the stress is too high, while the rest of the cable is still below that stress. This situation could easily be avoided by increasing the cross-sectional area at that point, leading to a larger heavier cable with more tension, but never sufficient stress to break.

There are many possible profiles in which the cable can be prevented from breaking. Here, we present one of the simplest: a cable for which the area varies such that the whole length is at, but never exceeds, the breaking stress.

This requires
\begin{equation}
T(\epsilon) = S \rho a(\epsilon) = \int dT
\end{equation}
which we can differentiate, using Equation~\ref{dTension}, to give the condition on the area
\begin{equation}
\frac{da}{d\epsilon} = \frac{a(\epsilon)}{\alpha} \left[\frac{1}{\epsilon^2} - \frac{\mu}{(1-\epsilon)^2} + (\mu - (1+\mu)\epsilon) \right].
\end{equation}
Integrating, using the boundary condition that $a=a_0$ at some $\epsilon=\epsilon_0$, we find
\begin{equation}
a(\epsilon) = a_0 e^{\frac{\eta(\epsilon) - \eta(\epsilon_0)}{\alpha}}.
\end{equation}

Finally, by definition
\begin{equation}
\sigma(\epsilon) = S.
\end{equation}

Notice that such a cable cannot possibly collapse ($T$ is always positive) nor break. However, there are still complications, hidden within this varying area.

Nothing prevents the area from becoming unphysically small, nor impractically large. Indeed, as $h$ gets smaller and smaller the maximum area (a useful proxy for the mass of the whole cable) grows as
\begin{equation}
a_{max} \approx a_0 e^{\frac{1}{h \alpha}}
\end{equation}
and thus for smaller $\alpha$ the mass of the cable must be huge in order to reach small $h$, near Earth's surface. %to near the Earth's surface.

This huge mass can be overcome by making $a_0$ very small, but there we hit other physical constraints; that at some point the cable becomes too small to produce and/or use. If we curtail the cable at some minimum area (and thus minimum $h$) this can be avoided, but in that case there is an even better solution, a hybrid cable.

\begin{table*}
\centering
\begin{tabular}{ c || c c c} 
  & $h < \epsilon < \epsilon_0$ & $\epsilon_0 < \epsilon < \epsilon_1$ & $\epsilon_1 < \epsilon < 1-\frac{r}{D}$ \\
 \hline
 \hline
 
 $a(\epsilon)$ & 
 $a_0$ & 
 $a_0 e^{\left(\frac{\eta(h)-\eta(\epsilon)}{\alpha}\right) - 1}$ & 
 $a_0$ \\
 
 $T(\epsilon)$ &
 $\frac{GM \rho a_0}{D}\left[\eta(h) -\eta(\epsilon) \right]$ &
 $\alpha \frac{GM \rho a_0}{D} e^{\left(\frac{\eta(h)-\eta(\epsilon)}{\alpha}\right) - 1}$ &
 $\frac{GM \rho a_0}{D}\left[\eta(h) -\eta(\epsilon) \right]$ \\
 
 $\sigma(\epsilon)$ & 
 $S \frac{\eta(h)-\eta(\epsilon)}{\alpha}$ & 
 $S$ & 
 $S \frac{\eta(h)-\eta(\epsilon)}{\alpha}$ \\
\end{tabular}
\caption{The area, tension and specific stress of a hybrid area cable, stretching from a height $h$ above the Earth's centre to an anchor point on the Moon. $\epsilon_0$ and $\epsilon_1$ give the positions where a cable of uniform area $a_0$ reaches its breaking stress (above and below the Lagrange point respectively). Note that for some combinations of $\alpha$ and $h$ this may not occur (see text).} 
\label{hybridTable}
\end{table*}

\begin{figure}[ht]
\includegraphics[width=\columnwidth]{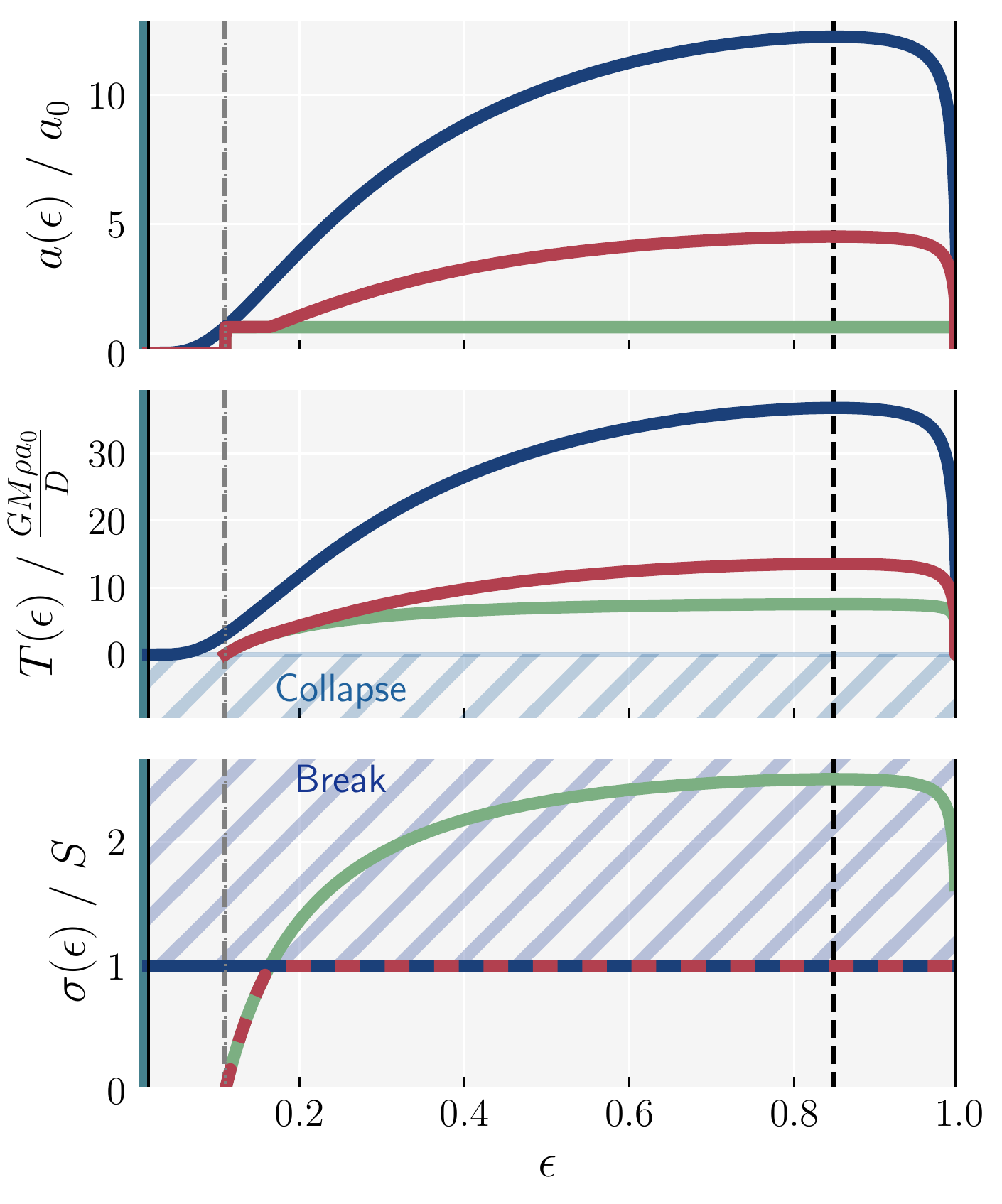}
\caption{The variation in area, tension and specific stress for a cable with $\alpha = 3$ extending from the Moon to geostationary orbit ($h = \epsilon_{\mathrm{geo}} = 0.12$). Solutions are shown for a uniform area cable (green), a tapered cable (blue), and a hybrid cable (red). Note that $\alpha = 3$ is too weak to allow for a uniform-area cable at this choice of $h$.}
\label{areaFig}
\end{figure}

\subsection{The hybrid cable - the most efficient practical solution}
\label{varying}

As we have shown, a varying-area cable can be constructed, for which we need not fear breakage or collapse, by enforcing that the cable be always at the maximum possible tension. But, if we choose a minimum cross-sectional area for such a cable, corresponding to a minimum height $h$, the free end is now thicker than it need be.

The most efficient solution is one in which we start at the Earth-end of the Spaceline with a constant area cable, as thin as is practical, which extends until the point at which it reaches its breaking stress, then tapers outwards from that point to avoid breaking. Past the Lagrange point, close to the Moon, where the tension (and therefore the allowable area) reduces again, there may be another section of uniform cable reaching down to the anchor point on the Moon's surface, though whether this second uniform-area section is possible depends on the value of $h$.

The area, tension and specific stress can be expressed relatively simply in terms of $h$. We define $\epsilon_0$ and $\epsilon_1$ as the points, above and below the Lagrange point, where the cable reaches its breaking stress and starts to taper. Thus they satisfy
\begin{equation}
    \eta(\epsilon_0) = \eta(\epsilon_1) = \eta(h) - \alpha
\end{equation}
where the exact values of $\epsilon_0$ and $\epsilon_1$ must be found numerically for particular choices of $\alpha$ and $h$. Then, we can find $a(\epsilon), T(\epsilon)$ and $\sigma(\epsilon)$, as shown in Table~\ref{hybridTable}.

Note that it is possible that $\epsilon_1 > 1-\frac{r}{D}$, in which case there is no uniform-area section between the Lagrange point and the Moon. This occurs when $\eta(h) \ge \eta(1-\frac{r}{D}) + \alpha$. For sufficiently high $\alpha$, or large $h$, the cable may never reach its breaking stress, and the most efficient solution is just that of a uniform-area cable.

This hybrid cable, by construction, cannot break but can collapse. In fact the same constraints (and solutions) apply here as did for the uniform-area cable. As long as $h$ is less than $\sim 0.24$, the cable will not collapse; for larger $h$, other solutions such as an anchorweight can similarly be implemented.

\subsection{Comparing profiles}

Figure \ref{areaFig} shows the area, tension and specific stress of the three cable designs for a cable with $\alpha=3$ extending to geostationary orbit ($h=0.12$). Note that the stress in the uniform area cable exceeds $S$, so a uniform-area cable of this strength would, in reality, break. It can clearly be seen how much switching to a hybrid cable reduces the area (and hence the mass) of the cable, and that a uniform area cable would be more efficient still except that it is not sufficiently strong given this choice of $\alpha$ and $h$.

\section{Practicalities and possibilities}
\label{practical}

We have shown that we have the materials and technology today that are necessary to support a spaceline construction. Now the remaining questions (of which there are many) relate to the practical constraints on such a construction and its potential uses.

\begin{figure}[ht]
\includegraphics[width=\columnwidth]{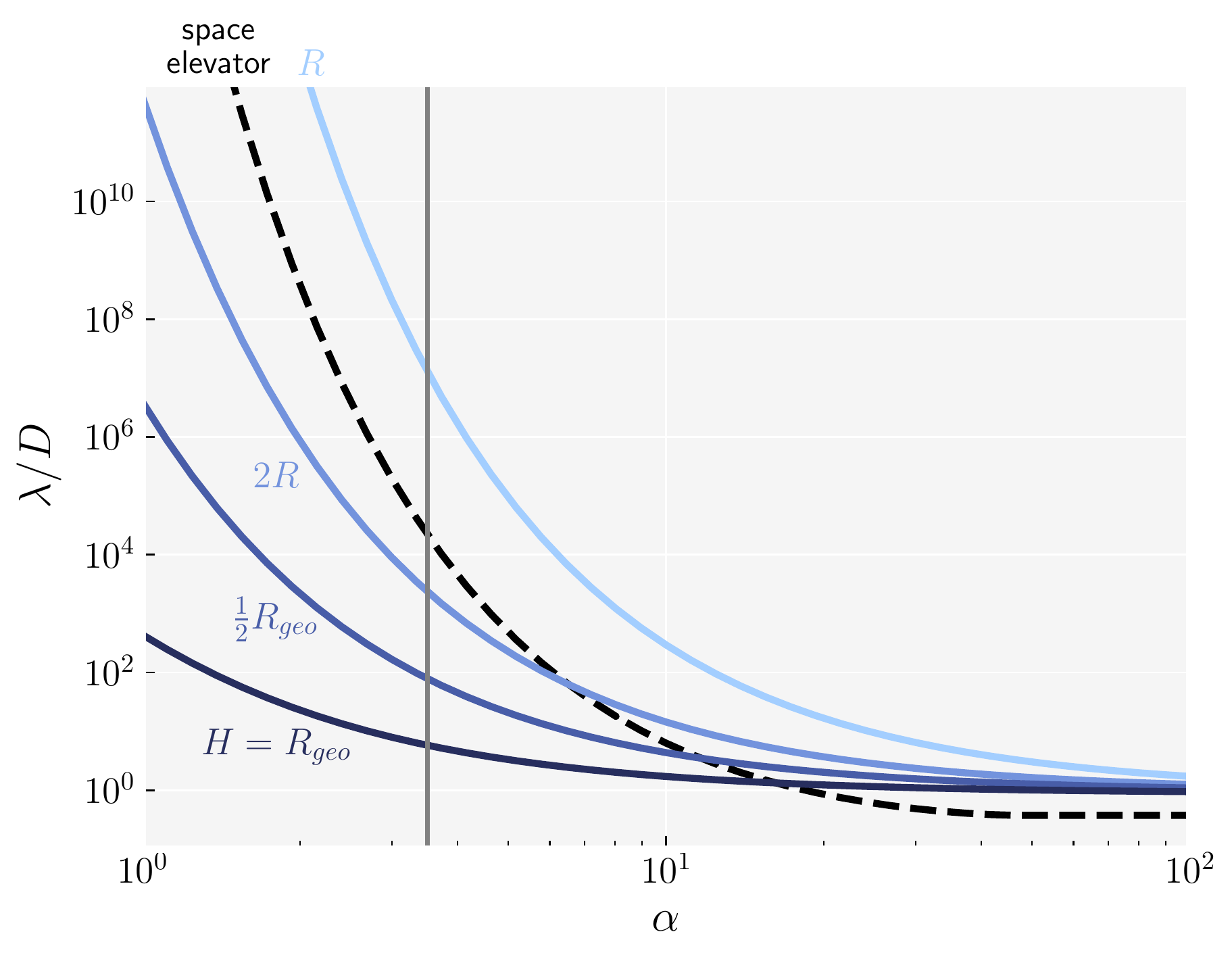}
\caption{The effective length of the spaceline as a function of the relative strength of the material. Different colours show the solution for different values of $H$. The dashed black line shows the effective length of the space elevator for comparison. Finally the grey line shows, for reference, the strength of the strongest mass-producible materials today.}
\label{lambdaFig}
\end{figure}

%\subsection{Mass of the spaceline}

The first question we must address is whether the mass of cable needed to construct a spaceline is feasible, given that it must be transported to and assembled in orbit.  

The mass of the cable is simply
\begin{equation}
m_{cable} = \int_h^{1-\frac{r}{D}} D a(\epsilon) \rho d\epsilon, 
\end{equation}
where the form of $a(\epsilon)$ will depend on the cable design.

Therefore the actual mass of the cable will depend on the chosen cross sectional area, $a_0$, which could vary wildly for different purposes, and on the height above Earth's centre $h$ that it extends to.

Thus it will be useful to measure the scale of the cable in another unit, the \textit{effective length},
\begin{equation}
\lambda = \frac{V_{cable}}{a_0} = \frac{m_{cable}}{\rho a_0} = D \int_h^{1-\frac{r}{D}} \frac{a(\epsilon)}{a_0} d\epsilon.
\label{equiv}
\end{equation}
This measure tells us how long a uniform-area cable of cross-sectional area $a_0$ would have to be to have total mass $m_{\mathrm{cable}}$. Thus, the effective length $\lambda$ can easily be converted to a mass for a given $a_0$ and $\rho$.

For uniform-area cables, this expression reduces to just $\lambda = 1 - h - \frac{r}{D}$, but for varying-area cables, $\lambda$ can become very large for small $h$. The lightest possible spaceline has an effective length of $\lambda \sim 0.75 D$. In comparison, the lightest possible space elevator has $\lambda \sim 0.4 D$ (though could not be supported by modern materials).

Figure \ref{lambdaFig} shows the effective length of the spaceline, and how it varies with $\alpha$ and $h$. We can see that for present-day materials a spaceline could be constructed which reaches geostationary orbit and has an effective length of order unity.

Let's say such a line was made of a cable with $a_0 = 10^{-7} m^2$: its total mass would then be around 40,000 kg. This is about twice the mass of the original lunar lander, and would make transporting and constructing such a cable completely plausible. The raw cost of the materials and transport could be numbered in the hundreds of millions of dollars.

Returning to Figure~\ref{lambdaFig}, we see that for cables which come significantly closer to Earth the effective length (and corresponding mass) increases rapidly. We also see that the spaceline can be much less massive, by many orders of magnitude, than a space elevator, although it would be more costly to build a spaceline extending all the way to Earth's surface than it would be to build a space elevator.

\subsection{Why build a spaceline}

This is not idle theorycrafting. Building a spaceline would be a huge engineering challenge, stretching the limits of current human capacity - but not exceeding them.

Even in its most economical form, a cable with a width only a little more than a pencil lead, it could cost billions of dollars for material and transport - and it is hard to quantify what extra cost such a project could incur. But a billion dollar price tag is not unattainable - and the possibilities of what could be done with such a structure may quickly pay dividends.

\begin{itemize}
\item Cost of transport - It costs slightly less, in fuel, to reach the spaceline than to move into a geostationary orbit (see appendix \ref{cost} for details). Transport along it is free - with solar powered climbing vehicles. This would reduce the cost of moving to anywhere along it's length substantially - for example it would reduce the fuel needed to reach the surface of the moon to a third of the current value.

\item Infrastructure - Objects in space float freely in a truly 3 dimensional space. But when you tether those objects to a line movement between them becomes a one dimensional journey. Motion between points on the spaceline is simpler and safer than moving, docking and navigating through empty space.

\item Haulage to and from the surface of the moon - Small loads can be sent to and from the moon. Whilst the moon is not the \textit{end destination} of the spaceline conceptually, it is physically. The ability to transport material to and from deep space without spaceflight changes the economics of scientific and industrial possibilities on the lunar surface.

\item The technological endeavour - An important but easy to overlook benefit of a project such as this is that the engineering challenges it presents, and how overcoming them will push forward our technological capacities. A difficult but achievable task requires us to make and master new techniques, and rewards us for doing so.

\item The Lagrange point base camp - The last item on this list is the thing we believe to be most important and influential for the early use of the spaceline (and for human space exploration in general). The spaceline makes the Earth Moon Lagrange point effectively stable. In this gravity free environment we can construct habitats and equipment of arbitrary mass (see appendix \ref{safety}). It is a pristine and gravity free environment, with no great hindrance to developing space constructions on a scale that would seem impossible otherwise. Having only a small team of scientists and engineers at such a base camp would allow hand construction and maintenance of a new generation of space based experiments - one could imagine telescopes, particle accelerators, gravitational wave detectors, vivariums, power generation and launch points for missions to the rest of the solar system. Equipment can be sent up in prefabricated pieces, assembled, maintained and operated by hand. Previous space stations have been constrained: by cramped conditions and dangerous environments. In contrast the Lagrange point is a haven for expansion and cultivation. It is the stepping stone from the surface of the Earth to stepping foot on the outer planets - and could be home to new populations, industry and enquiry on a scale yet unknown to us.
\end{itemize}

\subsection{Last words}

In this paper we hope to have presented a rational argument for the exploration of an idea that could pay huge dividends in it's scientific, economic and cultural impact. It is an idea much to large to be explored in anything but light detail in a single piece of work.

Many important questions remain unanswered, many technological and sociological challenges stand between the idea and it's execution. But if the logic presented here holds up to scrutiny \textit{it can be done}. With concerted effort and investment it may be a reality in decades, perhaps even years.

We hope it will inspire others to question, calculate, discuss, and to take a long view on our ambition and path to becoming a sustainable spacefaring civilisation. 
%\subsection{Cost of the space line}

%\subsection{Science enabled by the space line}

%\section{Discussion and conclusions}

%It works! Let's do it. Let's just build one. Go on, let's!

%WOoT!

\section*{Acknowledgements}

With thanks to Caleb Scharf, Cathie Clarke, Frits Paerels, Gillian Davis, Mike Payne, Nicholas C. Stone, Ryan MacDonald and many others, for pertinent questions and rational thoughts.

\appendix

\section{The classic space elevator}
\label{cElevator}

\begin{figure}
\includegraphics[width=10cm]{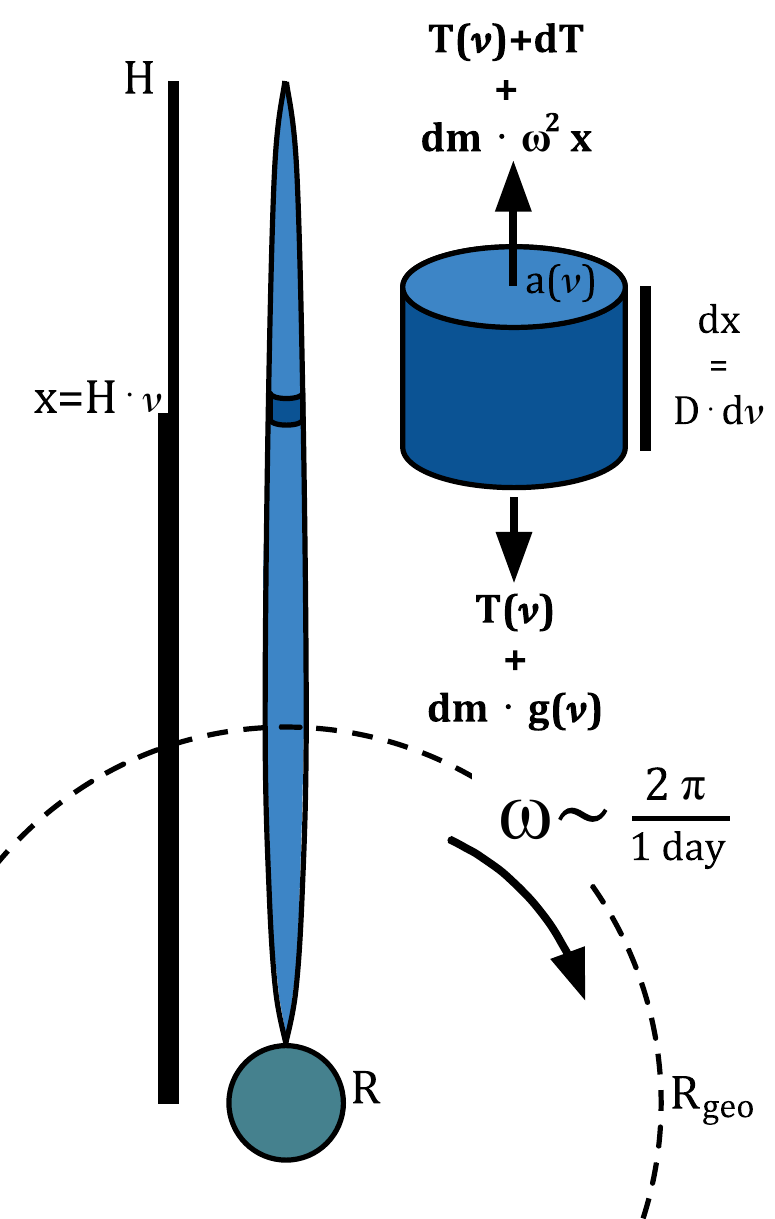}
\caption{A simple sketch of the space elevator, in a frame co-rotating with the Earth. Drawn to scale with the Earth and the radius of geostationary orbit.}
\label{elevator}
\end{figure}

In this section we will quickly re-derive the expressions similar expressions to those above, but specifically in relation to the classic space elevator. Readers may also want to refer to fuller works on the topic such as \citet{Aravind07} - here we simply map out the basic physics following the same formalism as applied to the Spaceline in the main body of the paper.

The space elevator extends from a anchor point on Earth's equator to above geostationary orbit. It is co-rotating with earth, with outward centrifugal force counteracting Earth's gravity. Figure \ref{elevator} shows a sketch of this system. 

The cable thus experiences an effective acceleration
\begin{equation}
g_{eff}(x) = \frac{GM}{x^2} - \omega^2 x
\end{equation}
where $\omega$ is the rotational velocity of Earth's surface, which (by definition) is equal to the orbital frequency at geostationary orbit:
\begin{equation}
\omega = \sqrt{\frac{GM}{R_{geo}^3}}.
\end{equation}

Using $\nu = \frac{x}{H}$, the effective force (in the corotating frame) on an infinitesimal mass element with mass $dm = \rho a dx$ is
\begin{equation}
d F_{eff} = \frac{GM \rho a}{R_{geo}} \left( \frac{1}{\nu^2} - \nu \right) d\nu.
\end{equation}

The tension in the cable can be found by integrating this from Earth's surface (at $\nu_0 = \frac{R}{H}$) giving
\begin{equation}
\label{tElevator}
T(\nu) = \frac{GM \rho}{R_{geo}} \int_{\nu_0}^{\nu} a(\nu') \left( \frac{1}{\nu'^2} - \nu' \right) d\nu'
\end{equation}
where we have set the tension at the anchor point on Earth to be zero, which minimises the overall tension in the cable.

\subsection{Constant area cable}

The simplest solution, much like for the spaceline, is a uniform area cable, with $a=a_0$.

This gives
\begin{equation}
\label{tUniform}
T(\nu) = \frac{GM\rho a_0}{R_{geo}} \left( \frac{1}{\nu_0} + \frac{\nu_0^2}{2} - \frac{1}{\nu} - \frac{\nu^2}{2} \right).
\end{equation}

The maximum tension in the cable occurs at geostationary orbit ($\nu=1$) and beyond this point the tension reduces as we move further from the Earth. Thus the condition for a uniform density cable not to break is $T(1) < T_{break}$. As above it will be convenient to work in terms of the \textit{relative strength} of the material compared to the strengths required for the problem:
\begin{equation}
\beta = \frac{R_{geo}S}{GM} = \frac{R_{geo}}{D} \alpha
\end{equation}
and thus the condition for the cable not to break is
\begin{equation}
\beta > \frac{1}{\nu_0} + \frac{\nu_0}{2} - \frac{3}{2} \approx 5.
\end{equation}

Given that $\beta \approx \frac{\alpha}{10}$ this means that no mass producible materials are strong enough to construct such a space elevator, though carbon nanotubes would be (see table \ref{materials} for reference).

The length of the cable could be reduced by a counterweight above geostationary orbit, and these are commonly shown in space elevator design. However such a weight will increase the mass of the total construction, whilst reducing how high the elevator can reach, and so we do not consider it further here.

\subsection{Varying area cable}

Again, as we show in Section \ref{varying} for the spaceline, we can make an optimal space elevator by varying the area of the cable such that the whole length is at (or just below) the point of breaking.

From equation \ref{tElevator} we can see that by setting the tension equal to $T_{break} = S \rho a$ and differentiating we find
\begin{equation}
\label{aElevator}
\beta \frac{da}{d\nu} = a(\nu) \left( \frac{1}{\nu^2} - \nu \right)
\end{equation}
and integrating this, letting $a=a_0$ at $\nu=\nu_0$ gives
\begin{equation}
a(\nu) = a_0 e^{\frac{1}{\beta}\left( \frac{1}{\nu_0} + \frac{\nu_0^2}{2} - \frac{1}{\nu} - \frac{\nu^2}{2} \right)}.
\end{equation}

In theory we could build a cable such as this with any material, but for weak materials with small $\beta$ the area (and hence the mass of cable) becomes prohibitively large at geostationary orbit.

\subsection{The optimal hybrid cable}

If we seek to minimise the mass of a cable of a given material the optimal solution is a hybrid cable. This takes into account there being some minimum producible and useful area. 

Such a cable starts with a length of some minimum area $a_0$ extending up to some $\nu_1$ where the tension becomes so high the cable would break. From this point it tapers outwards, keeping the stress in the cable just below breaking point. Beyond geostationary orbit the area starts to decrease, and then when it again reaches the minimum possible area ($a=a_0$) at $\nu_2$ we revert to a constant area cable with sufficient length to act as the counterweight.

From equation \ref{tUniform} we see that $T(\nu_1) = T_{break}$ when
\begin{equation}
\beta = \frac{1}{\nu_0} + \frac{\nu_0^2}{2} - \frac{1}{\nu_1} - \frac{\nu_1^2}{2}.
\end{equation}
Between $\nu_1$ and $\nu_2$ the area still obeys equation \ref{aElevator} and integrating setting $\nu=\nu_1$ when $a=a_0$ gives
\begin{equation}
\label{eHybrid}
a(\nu) = a_0 e^{\frac{1}{\beta}\left( \frac{1}{\nu_0} + \frac{\nu_0^2}{2} - \frac{1}{\nu} - \frac{\nu^2}{2} \right) - 1}.
\end{equation}

Setting equation \ref{eHybrid} equal to $a_0$ and solving for $\nu$ yields a cubic which can be solved for $\nu_2 (>1)$. Beyond this point the tension follows
\begin{equation}
T(\nu) = \frac{GM\rho a_0}{R_{geo}} \left(\beta + \frac{1}{\nu_2} + \frac{\nu_2^2}{2} - \frac{1}{\nu} - \frac{\nu^2}{2} \right)
\end{equation}
and at some $\nu_3 > \nu_2$ the cable terminates, at the point at which $T$ goes to zero.

For sufficiently strong materials ($\beta > \sim 5$ or equivalently $\alpha > \sim 50$) the cable may never reach its breaking tension and hence $a=a_0$ throughout.

Summarising
\begin{equation}
\begin{aligned}
a(\nu) \ = \ &a_0 &\nu_0 < \nu < \nu_1 \ \mathrm{where} \ T(\nu_1)=T_{break} \\
&a_0 e^{\frac{1}{\beta}\left( \frac{1}{\nu_0} + \frac{\nu_0^2}{2} - \frac{1}{\nu} - \frac{\nu^2}{2} \right) - 1}  &\nu_1 < \nu < \nu_2 \ \mathrm{where} \ T(\nu_2) = T_{break} \\
&a_0 &\nu_2 < \nu < \nu_3 \ \mathrm{where} \ T(\nu_3) = 0 \\
&0  &\nu > \nu_3.  \\
\end{aligned}
\end{equation}

Figure \ref{lambdaFig} shows the effective length (see equation \ref{equiv}, the length of an equivalent mass cable of constant cross sectional area) needed to construct a space elevator. The hybrid cable reduces the weight by a factor of $\sim 4$ compared to a varying area cable, and has the same mass as a uniform area cable (not shown) for sufficiently strong materials.

\section{Transfer orbits}
\label{cost}

One of the most immediate appeals of constructing a spaceline is the direct saving in fuel cost for missions going to the Lagrange point or the moon. In this section we will map out some simple calculations of the positive impact of using the spaceline.

We will talk about cost here not in terms of a monetary value, but in mass of fuel. There are of course other costs involved in any space bound mission, which may dwarf fuel costs, but those are beyond the scope of simple calculation. All we can say with relevance to these costs is that they will inevitably reduce as space travel becomes more prevalent, whilst the fuel cost gives a hard limit on the economy of space travel.

One crucial point to remember about the spaceline is that it is traversable without any energy cost. A solar-powered climber, gripping the line with two wheels (in the simplest design) can freely move up and down the thread without any cost in fuel.

Thus the cost of getting to all point on the spaceline is constant, except for the time it takes. Travel to and from the moon down to geostationary orbit is effectively free.

Fuel costs can usefully be translated into the convenient physical unit, $\Delta v$, the total change in velocity needed to reach a certain orbit. This in turn is really a measure of the energy difference between one orbit and another, as we trade kinetic energy for gravitational potential.

This can then be related to fuel costs via the Tsiolkovsky rocket equation:
\begin{equation}
\Delta v = v_e \ln{\frac{m}{m_0}}
\label{rocket}
\end{equation}
where $m$ is the current mass of the spaceship, $m_0$ is the initial mass and $v_e$ is the exhaust velocity of the engines, the speed at which each engine can expel mass.

Thus the mass of fuel needed to achieve a certain $\Delta v$ is
\begin{equation}
m_F = m_0 - m = m \left(e^\frac{\Delta v}{v_e} - 1 \right).
\end{equation}

\subsection{Orbital manoeuvres}

We will compare two orbital manoeuvres here, a Hohmann transfer orbit, taking a payload from Earth to a circular orbit, and a spaceline transfer, bringing the same payload to the spaceline at the appropriate velocity to dock.

Both of these will be achieved by an impulsive (meaning approximately instantaneous) fuel burn at Earth, putting the ship onto an eccentric orbit with apoapse at the desired distance from Earth. The difference between the two orbits is seen at apoapse:
\begin{itemize}
\item Circular transfer: The ship is moving well below the circular velocity at apoapse, so we need to make another impulsive burn to match the circular velocity.
\item Spaceline transfer: The spaceline is moving at a speed set by the moons angular velocity, which the ship must match with an impulsive burn. This may be an acceleration or deceleration depending on the relative velocity which depends on distance from Earth.
\end{itemize}

To calculate the changes of velocity we will make extensive use of the Vis-Viva equation, relating the current velocity ($v$) and radius ($r$) of the orbit to the semi-major axis ($a$):
\begin{equation}
v^2 = GM\left[\frac{2}{r} - \frac{1}{a} \right]
\label{visviva}
\end{equation}
where $M$ is the mass of the Earth (we will ignore the mass of the moon for these calculations presently).

We will also use the fact that for an eccentric orbit the relationship between the apo and periapse radii ($r_\pm$), semi-major axis and eccentricity ($e$) is
\begin{equation}
r_\pm = a (1 \pm e).
\label{apse}
\end{equation}

\subsection{Leaving Earth}

We will consider orbits starting at the surface of the earth, i.e. $r_- = R$, and reaching a distance from earth $r_+ = \alpha R$. This allows us to calculate the energy cost in the general case and then relate it back to specific distances from Earth. For example geostationary orbit has $\alpha \approx 6.6$, the Earth-Moon Lagrange point is at $\alpha \approx 51$ and the Moon itself is at $\alpha \approx 60$.

We can rearrange equation \ref{apse} to find
\begin{equation}
a =\frac{\alpha + 1}{2} R \ \ \mathrm{and} \ \ e=\frac{\alpha-1}{\alpha+1}.
\end{equation}

Thus from equation \ref{visviva} the velocity needed at periapse ($r_- = R$) for the intital transfer, an ellipse extending the $\alpha R$, is
\begin{equation}
v_- = v_0 \sqrt{\frac{2\alpha}{\alpha+1}}
\end{equation}
where
\begin{equation}
v_0 = \sqrt{\frac{GM}{R}} \approx 7.9 kms^{-1}
\end{equation}
is a characteristic velocity of orbital maneuvers around Earth.

This ignores the effort of leaving Earth's atmosphere (a near negligible difference in gravitational potential but significant work done against air resistance). We will encapsulate this factor in a simple constant and leave further analysis for other works. Let $\beta$ be the approximate $\Delta v$ spent on drag forces in the atmosphere, and we will use a value of $\beta = 1.5 kms^{-1}$ for the sake of calculations.

The rotational velocity of the Earth at the Equator is $v_E \approx 0.4 kms^{-1}$ and hence to leave Earth we need $(\Delta v)_E \approx v_- - v_E + \beta$.

We can also use equation \ref{visviva} to find the velocity at apoapse ($r_+ = \alpha R$):
\begin{equation}
v_+ = v_0 \sqrt{\frac{2}{\alpha(\alpha +1)}}.
\end{equation}

\subsection{Circular transfers}

Returning once more to equation \ref{visviva} we can find the velocity needed for a circular orbit at $r= a =\alpha R$:
\begin{equation}
v_c = v_0 \sqrt{\frac{1}{\alpha}}.
\end{equation}

This is always greater than $v_+$ and hence for a ship to be able to maintain this altitude it must perform another burn with $(\Delta v)_+ = v_c - v_+$.

Putting this all together we find the total $\Delta v$ needed for a stable orbit at $r=\alpha R$:
\begin{equation}
(\Delta v)_c = v_0 \left[\sqrt{\frac{2\alpha}{\alpha+1}} + \sqrt{\frac{1}{\alpha}} - \sqrt{\frac{2}{\alpha(\alpha+1)}} \right] + \beta - v_E
%v_0 \sqrt{\frac{2}{\alpha(\alpha+1)}} \left[\alpha + \sqrt{\frac{\alpha+1}{2}} - 1 \right] + \beta - v_E
\label{deltaCircular}
\end{equation}
which tends to $\sim \sqrt{2} v_0 + \beta - v_E$ for large $\alpha$.

\subsection{Transfers to the spaceline}

The velocity of the spaceline at a given radius is set by the orbital velocity of the moon, which has angular velocity
\begin{equation}
\omega_m = \sqrt{\frac{GM}{R_m^3}} = \mu^{-\frac{3}{2}} \frac{v_0}{R}
\end{equation}
where $\mu \approx 60$.

Thus the velocity of the line at a distance $\alpha R$ from Earth is
\begin{equation}
v_l = \alpha R \omega_m = \frac{\alpha}{\mu^{\frac{3}{2}}} v_0.
\end{equation}

At larger distances we can see that $v_l$ increases whilst $v_+$ decreases, hence beyond some distance from Earth the line transitions from moving slower than the ship to overtaking it. Thus the change in velocity needed at apoapse to meet the spaceline is $|v_l - v_+|$.

This gives a total $\Delta v$ needed to dock with the spaceline of
\begin{equation}
(\Delta v)_l = v_0 \left[\sqrt{\frac{2\alpha}{\alpha+1}} + \left|\frac{\alpha}{\mu^{\frac{3}{2}}} - \sqrt{\frac{2}{\alpha(\alpha+1)}}\right| \right] + \beta - v_E.
\label{deltaLine}
\end{equation}

\subsection{Comparing transfer manoeuvres}

\begin{figure}
\includegraphics[width=10cm]{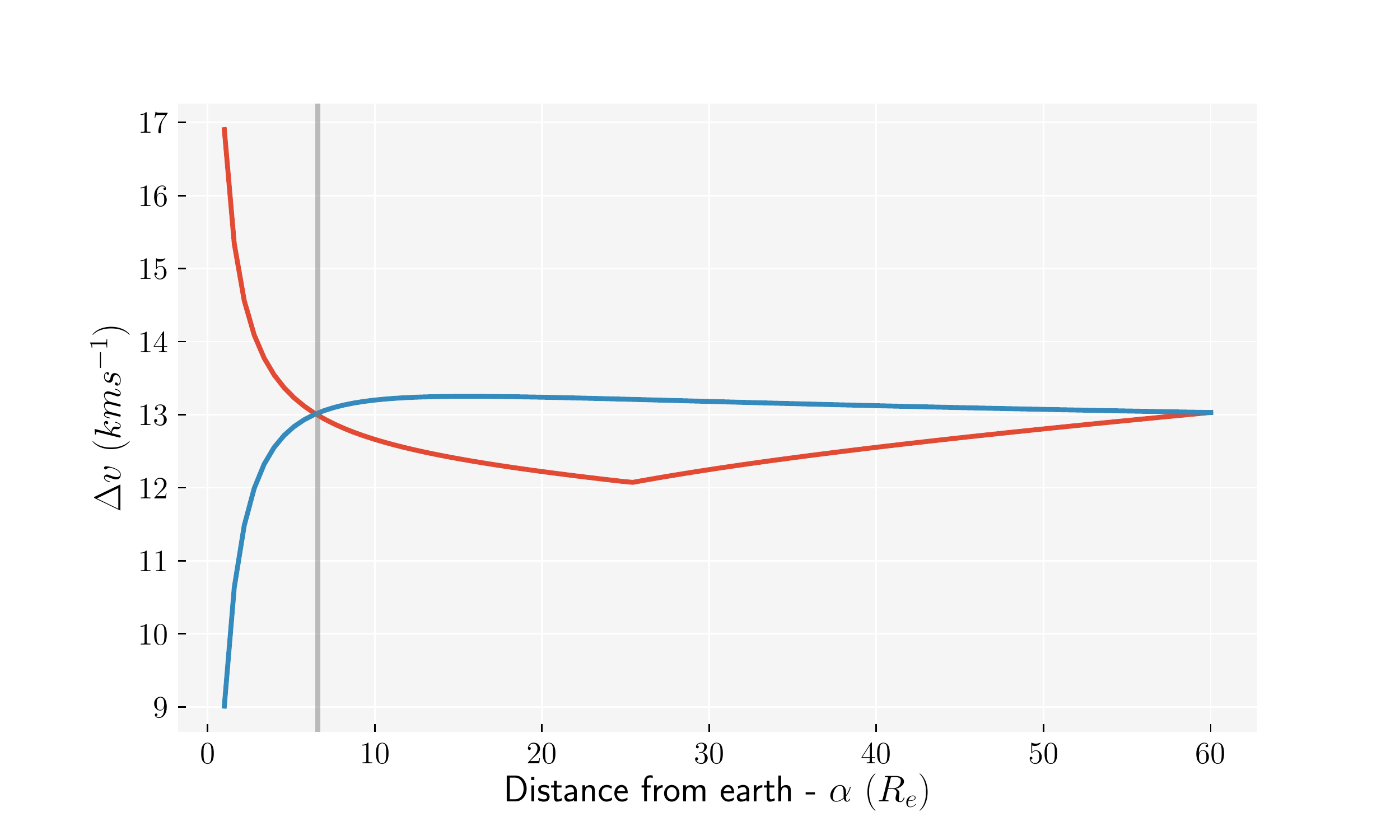}
\caption{$\Delta v$ needed to go from Earth to a distance $\alpha R$. The cost of a transfer to a circular orbit is shown in blue, and for docking with the spaceline in red. The vertical line shows the location of geostationary orbit (I believe it's pure coincidence that this is also very near where the tow lines meet).}
\label{deltaCompare}
\end{figure}

We can now compare the $\Delta v$ needed for a transfer to a circular orbit (equation \ref{deltaCircular}) and to docking with the spaceline (equation \ref{deltaLine}). This is shown in figure \ref{deltaCompare}, where we can see that transferring to the spaceline is a significantly cheaper way to get to any orbital radius beyond $\alpha \approx 3$. Note that the $\Delta v$ required to actually approach the moon directly is significantly higher than shown here, as a second set of orbital manoeuvres is needed to slow down enough to orbit and to approach the surface. Approaching the moon via the spaceline has no extra cost.

Finally we can look at the direct cost in fuel of these two maneouvers using equation \ref{rocket} and a typical exhaust velocity of $v_e = 3.5 kms^{-1}$. We see that using the spaceline can save around 25\% of the cost per launch in fuel if we go to the most efficient point (at $\alpha \approx 25$) and then use the line itself to traverse the remaining distance for free. Even meeting it at geostationary orbit height it is as efficient as a normal circular geostationary orbit. 

Going to the Lagrange point is a similar saving (and much less operationally complex than having to navigate a region that may soon be filled with scientific equipment).

The comparative cost of going to the moon is not well represented here, as explained previously. But using some rough $\Delta v$ estimates we can see that the spaceline cuts the fuel cost of such a mission by approximately two thirds.

\begin{figure}
\includegraphics[width=10cm]{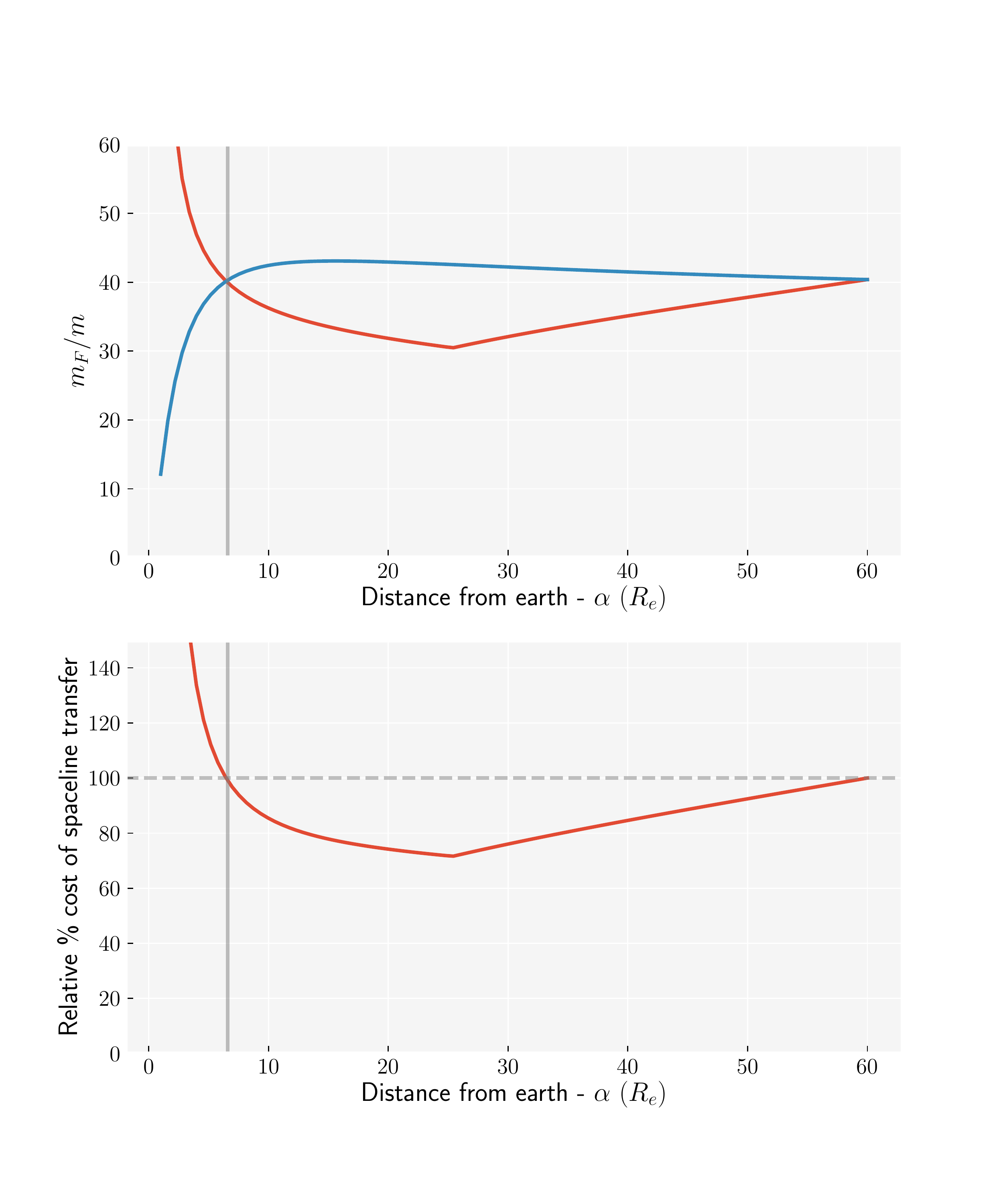}
\caption{Top plot: Mass of fuel, relative to payload, needed for a circular orbit (blue) compared to docking with the spaceline (red) using an exhaust velocity of $v_e = 3.5 kms^{-1}$. Bottom plot: Fuel cost of the spaceline transfer compared to circular orbit.}
\label{massFuel}
\end{figure}

\section{The safeline}
\label{safety}

We have calculated the tension and stresses on the spaceline, and showed that with modern materials it could be constructed within the fundamental limits of the materials. Now we can extend those calculations to explore the practical limits, the compromises that can be made to ensure the safe use of a spaceline and the constraints they cause.

\subsection{A simple safety factor}

There are two major events we wish to avoid with the spaceline:
\begin{itemize}
\item Breaking - if the stress at any point in the cable exceeds the breaking strength the line will snap. The highest tension is felt at the Lagrange point, and is a function of how deep into Earth's potential well the cable extends.
\item Collapsing - if the tension in the cable anywhere $<0$ the cable cannot support itself (it is only strong in tension). The only place where negative tension can occur is near to the surface of the moon, and this can be avoided by a longer cable which feels a greater force throughout.
\end{itemize}

These two conditions must be balanced to build a usable cable, giving a range of possible cable lengths. If the cable is allowed to vary in area a wider range of solutions exist, but for simplicity we will not consider them here. For materials below a certain strength cutoff (roughly that of common carbon fibre) there are no possible cable heights for which the cable can be constructed.

Materials such as Kevlar and Dyneema are strong enough to allow a range of solutions. We can define the relative strength of the material, the ration of the specific strength to the typical stresses involved in the Earth-Moon system -
\begin{equation}
\alpha = \frac{S D}{G M}.
\end{equation} The specific stress in the cable follows
\begin{equation}
\sigma(\epsilon) = \frac{S}{\alpha}\left[\eta(h) - \eta(\epsilon) \right]
\end{equation}
where
\begin{equation}
\eta(\epsilon) = \frac{1}{\epsilon} + \frac{\mu}{1-\epsilon} +\frac{(1+\mu)\epsilon^2}{2} - \mu \epsilon.
\end{equation}
Here $\epsilon(=x/D)$ is a scaled dimensionless parameter giving the distance from the Earth, $\mu (=m/M)$ is the dimensionless ratio of lunar mass to Earth's and $S$ is the breaking stress.

The cable breaks when $\sigma(l) \ge S$ (where $l \approx 0.9$ is the scaled distance to the Lagrange point) and collapses if $\sigma(1-\frac{r}{D}) \le 0$ (where $1-\frac{r}{D}$ is the scaled distance to the surface of the Moon).

Let's define the excess specific stress that can be applied before breaking:
\begin{equation}
\sigma_b = S -\sigma(l) = \frac{G M}{D}\left[\alpha + 1.6 - \eta(h)\right]
\end{equation}
and the limiting specific stress that could be removed before the cable collapses:
\begin{equation}
\sigma_c = \sigma\left(1 - \frac{r}{D}\right) = \frac{G M}{D} \left[\eta(h) - 3.5\right]
\end{equation}
(using $\eta(l) \approx 1.6$ and $\eta(1-\frac{r}{D}) \approx 3.5$).

We can use the useful approximate result that far from $h=0$ or $h=1$, $\eta(h) \approx \frac{1}{h}$ and thus for a cable with area $a_0$ and density $\rho$ the tension that can be added to the cable is
\begin{equation}
T_b = \frac{G M}{D} \rho a_0 (\alpha + 1.6 -\frac{1}{h}).
\end{equation}
Similarly the tension that can be taken away from the cable before it collapses is
\begin{equation}
T_c = \frac{G M}{D} \rho a_0 (\frac{1}{h} - 3.5).
\end{equation}

Thus we can see the limit on such a material occurs for $T_c = T_b = 0$ where $\alpha \approx 0.9$.

To give an example using a specific candidate material for the spaceline, when $\alpha = 3.5$, roughly the value for the stronger variants of Dyneema (which also has $\rho = 970 kg m^{-3}$ and also using $\frac{G M}{D} \approx 10^6$) we find $T_b = 10^9 a_0 (5.1 - \frac{1}{h}) N m^{-2}$ and $T_c = 10^9 a_0 (\frac{1}{h} - 3.5)$. Thus for $h \approx 0.25$ and a cable of area $10^{-7} m^2$ (a suggested feasible minimal area for the first iteration of the spaceline) $T_c \approx T_b \approx 100 N$.

Another way of thinking about the above calculation is that it gives the loading weight that can be applied at the Earth (Moon) end before the cable breaks (collapses). This means we could load over $2000 kg$ at the Earth end of the cable, or over $6000 kg$ at the most fuel-efficient point to meet the spaceline ($h \approx 0.4$, see section \ref{transfer}). At the moon (where gravity is almost exactly one-tenth that on Earth, $\approx 1 N kg^{-1}$) this would only allow transport of weights up to $100 kg$. These numbers rise quickly for stronger materials (larger $\alpha$) and thicker cables (larger $a_0$), scaling approximately linearly with both.

This may suggest that the first proposed spaceline, with an area of $10^{-7} m^2$ is of more direct relevance and use for deep-space work, and subsequent construction (surpassing $a_0 > 10^{-6}$) will be of more interest for Lunar work. It is also worth pointing out that this is an approximate calculation and a factor of two or more weight may be transportable with carefully tweaked parameters.

\subsection{Lagrange base-camp}

Building a \textit{base-camp} at the Lagrange\footnote{Often termed the L1 point, one of the five stationary points in the restricted two-body problem} point is one of the most immediately useful and exciting utilities of the spaceline. A small habitat there could house many scientists and engineers, much like the Antarctic base camp. This would allow experimentation and construction in a near-pristine, gravity-free environment. 

There are two huge advantages of fabricating and assembling structures at the Lagrange point rather than any other stable orbit:
\begin{itemize}
\item No debris - The region of space between Earth and geostationary orbit is filled with the remnants of past missions and abandoned satellites. Also, stable (and thus long-lived) fast moving orbits can exist here, raising the fear of bombardment with naturally occurring meteoroids. The Lagrange point has been mostly untouched by previous missions, and orbits passing through here are chaotic, greatly reducing the amount of meteoroids.
\item Non-dispersive - If you drop a tool from the ISS it will seem to rapidly accelerate away from you. This is because of the slight difference in the gravitational force felt at different distances from the Earth, leading to orbits that quickly diverge. This makes it a difficult and dangerous place for construction. The Lagrange point has an almost negligible gradient in gravitational force, the dropped tool will stay close at hand for a much longer period. With small corrective thrusters or a minimal system of tethers, many objects (habitats, science equipment or spacecraft) can be held in a stable configuration indefinitely. Space now has a "next-door".
\end{itemize}

Manned large-scale construction projects would become much easier to build and maintain. These could include a new generation of significantly larger space telescopes, a network of isolated gravitational wave detectors and particle accelerators on scales much surpassing what can feasibly be built upon Earth's surface.

Similarly, the base camp itself can be extended, with prefabricated panels added to allow increased space for habitation and experimentation. Scientific and industrial testing in vacuum or zero-gravity environments can be undertaken over longer periods and bigger scales than previously imaginable.

There is one caveat though, the nature of the Lagrange point between the Earth is unstable. The effective potential (in the corotating frame) is a saddle point. If an object undergoes small displacements in the tangential direction (constant radius) the will feel a restoring force back to the Lagrange point. However, if the object wanders in the radial direction (towards the Moon or Earth) it will be pulled more and more strongly in that direction. Thus to keep an object at the Lagrange point indefinitely there needs to be a corrective force in the radial direction.

The spaceline naturally provides this force, and this is one of the two major reasons why constructing a spaceline makes a Lagrange point base camp significantly easier to use and maintain. The other being that it allows material transport easily to and from the base camp (via a spaceship carrying material from Earth, or directly from the surface of the moon), without the need for coordinating rocket flight through a region of space that may quickly fill with delicate habitats and scientific equipment.

In the simplest version of the \textit{safeline} there can be a force of up to $100 N$ either towards Earth or the Moon before there is any danger of the cable breaking or collapsing.

The acceleration felt by a mass near the Lagrange point along the radial direction is approximately
\begin{equation}
a = 10 \frac{G M}{D^3} x
\end{equation}
where $x$ is the distance from the Lagrange point. Thus for a restoring force from the cable of up to $100 N$ and a base camp with $10^6 kg$ mass, it can wander up to $\sim 400 km$ from the exact Lagrange point. Thus it is relatively simple and safe to ensure the base camp stays a stably near to the Lagrange point.

\subsection{Other safety concerns}

There are two more points that we have considered with regards to the safety of the cable - but are beyond the scope of this paper to address here: stability and impacts.

\subsubsection{Stability}

Earth's gravity anchors the end of the spaceline, always pulling it back towards straight. However the system is in a rotating frame, and as we begin to have substantial movement of mass along the cable we will generate motion, via the Coriolis force.

There is no inherent damping in the systems, and due to the varying tension along the length of the cable the propagation of waves along it's length is not trivial to calculate.

More in depth analysis will be needed to assess whether introducing motion to the system could lead to an unstable state, and how best this can be addressed. Depending on the magnitude of energy input steps could be taken to damp the motion, ranging from increasing the natural damping of the cable, to solar sails or corrective thrusters.

\subsubsection{Impacts}

Close to gravitating bodies micrometeroids will accumulate. Though they may be almost imperceptibly small, they could still damage or even break the cable upon impact.

The simple solution to this is to distribute the tension in the cable across multiple strands, such that one or more can break without greatly reducing the strength of the cable.

These broken strands could theoretically then be repaired systematically, much like small damages to a railway line.

The problem can be further contained by breaking the cable up into individual spans - many strands all connected to a terminating plate at each end - such that a breakage of one strand only affects the strength of that span, not the cable as a whole.

To fully understand the measures that must be taken to reduce this risk the rate at which such impacts might occur must be calculated, which is beyond the scope of this paper.

\bibliography{bib}

\end{document}